\documentclass[%
aip,
graphicx,
amsmath,amssymb,
reprint
]{revtex4-1}

\draft 
\usepackage{graphicx}
\usepackage{dcolumn}
\usepackage{bm}

\usepackage[utf8]{inputenc}
\usepackage[T1]{fontenc}
\usepackage{mathptmx}
\usepackage{etoolbox}
\makeatother
\newcommand{\bbr}{\boldsymbol{r}}
\begin{document}


\title{Variational principle to regularize machine-learned density functionals: the non-interacting kinetic-energy functional} 



\author{Pablo del Mazo-Sevillano}
\email[]{pablo.delmazo@uam.es}
\affiliation{Universidad Autónoma de Madrid, Departamento de Química Física Aplicada, Módulo 14, 28049 Madrid, Spain}
\affiliation{FU Berlin, Department of Mathematics and Computer Science, Arnimallee 12, 14195 Berlin, Germany}
\author{Jan Hermann}
\affiliation{FU Berlin, Department of Mathematics and Computer Science, Arnimallee 12, 14195 Berlin, Germany}
\affiliation{Microsoft Research AI4Science, Karl-Liebknecht-Str. 32, 10178 Berlin, Germany}


\date{\today}

\begin{abstract}
Practical density functional theory (DFT) owes its success to the groundbreaking work of Kohn and Sham that introduced the exact calculation of the non-interacting kinetic energy of the electrons using an auxiliary mean-field system. However, the full power of DFT will not be unleashed until the exact relationship between the electron density and the non-interacting kinetic energy is found. Various attempts have been made to approximate this functional, similar to the exchange--correlation functional, with much less success due to the larger contribution of kinetic energy and its more non-local nature. In this work we propose a new and efficient regularization method to train density functionals based on deep neural networks, with particular interest in the kinetic-energy functional. The method is tested on (effectively) one-dimensional systems, including the hydrogen chain, non-interacting electrons, and atoms of the first two periods, with excellent results. For the atomic systems, the generalizability of the regularization method is demonstrated by training also an exchange--correlation functional, and the contrasting nature of the two functionals is discussed from a machine-learning perspective.
\end{abstract}


\maketitle 

\section{\label{sec:intro} Introduction}
Density functional theory (DFT) has become an essential tool of every computational chemist and condensed-matter physicist thanks to its ability to accurately predict the electronic properties of molecules\cite{mardirossian:2017} and materials \cite{cramer:2009, jain:2016} at an accessible cost. DFT is a quantum-mechanical approach that provides a practical and efficient way to obtain the electronic structure of molecules, solids, and surfaces.

There are two terms to approximate in DFT as formulated by \citet{hohenberg:1964} (HK): the kinetic energy and the exchange--correlation (XC) energy. In 1965, Kohn and Sham\cite{kohn:1965} (KS) proposed a way to avoid drastically approximating the kinetic energy by introducing a set of one-electron orbitals that allow for exact calculation of the non-interacting kinetic energy, and including the small difference to the exact kinetic energy in the XC term. As a result, only the XC functional remains to be approximated. Despite its crucial role in capturing the chemistry of molecules and materials, crude approximations of the XC functional are often sufficient as it typically accounts for a small fraction of the total energy \cite{wang:2002}.
The core principle of DFT is the fact that only the electron density, a 3-dimensional function, is required to fully describe the electronic structure and energetics of a system. While KS-DFT is extremely useful, it deviates from this fundamental principle by requiring the use of a set of KS orbitals.

While the development of newer and more accurate XC functionals has received a wealth of attention, the same level of focus has not been directed towards kinetic-energy functional (KEF), which enables the orbital-free DFT (OF-DFT). The reason for this could simply be pragmatic, as KS-DFT provides the exact non-interacting kinetic energy, together with the increasing difficulty of producing accurate functionals given its non-local nature and its large contribution to the total electron energy. Nonetheless, the lack of access to precise non-interacting KEF limits our ability to fully utilize the computational advantages of HK-DFT.

The non-interacting KEF can be conveniently and exactly partitioned into two components\cite{sears:1980, deb:83}: the von Weizsäcker and Pauli functional. 
The former is the only contribution for fermionic systems described with a single orbital or for bosonic systems, while the latter embodies the effects of the Pauli principle (antisymmetry) \cite{march:86, levy:88}. Although finding a suitable approximation of the Pauli functional has proven to be a formidable task, some of its exact properties can be derived, such as its positiveness and scaling conditions \cite{levy:88}. Many attempts have been made to produce accurate Pauli functionals, from semi-local approximations that do not usually yield accurate results, certainly not without incorporating at least the Laplacian of the density \cite{perdew:2007, francisco2021}, to more sophisticated non-local functionals \cite{chacon:1985, garcia:1996, garaldea:2008, sarcinella:2021}. The former are defined in real space and fail to reproduce the Lindhard function and the latter are defined in reciprocal space, which complicates their use for finite systems \cite{sarcinella:2021}.

Machine learning (ML) is an excellent tool for unravelling complex mappings between a set of inputs and outputs that are otherwise difficult to formulate theoretically. As a result it has permeated almost every branch of science\cite{jumper:2021, ravuri:2021, Lin:2022, degrave:2022}.
Electronic structure calculations have benefited from this development---for instance variational quantum Monte Carlo from more flexible \textit{ansatzes} \cite{choo:2020, pfau:2020, hermann:2020a} and DFT in form of ML XC functionals \cite{dick:2020,kirkpatrick:2021,kasim:2021}.

One path to improve the density functionals is to make use of more complex analytical forms satisfying more exact constraints\cite{sun:2015} and climbing the Jacob's ladder \cite{perdew:2001}, but there has been also criticism that some of the more recent functionals are biased towards computing correct energies, rather than producing both correct energies and densities \cite{medvedev:2017}. ML together with large amounts of electron densities and energies may help in the search of more accurate and unbiased density functionals. The pioneering work in this field comes from J. C. Snyder \textit{et al.} \cite{snyder:2012} who, for the first time, proposed to learn a density functional ---the KEF for a set of non-interacting electrons. One of the main conclusions of that work was that it is not possible to only train very flexible functionals by matching the energies to some reference data, but also its functional derivative must be regularized in order to obtain a functional that is applicable in a self-consistent calculation. Since this work, several regularization methods have been developed, such as the use of the KS equations as regularizers, from L. Li \textit{et al.} \cite{li:2021}, or minimizing the second-order change of the energy from a single SCF step from J. Kirkpatrick \textit{et al.} \cite{kirkpatrick:2021}, both of which have been applied to learning the XC functional.

In this work we propose a new regularization method focused on the training of the KEF which is derived from the variational principle, imposing that the electron energy functional has a minimum at the reference density. The paper is organized as follows. Section~\ref{sec:theory} presents a short introduction to DFT with a derivation of the minimum condition for the energy functional needed in our regularization. We then introduce the gradient regularization, after discussing previous methods, arguing for the need of this new form. In Section~\ref{sec:results} we evaluate the effectiveness of gradient regularization by training the KEF for three (effective) 1D systems, namely the hydrogen chain, non-interacting electrons, and atoms. We demonstrate its broad applicability by using it to train the XC functional for a group of atoms and comparing the training methodologies for both the KEF and XC functional. Finally Section~\ref{sec:conclusions} summarizes the results and presents future perspectives in density-functional learning.

\section{\label{sec:theory} Theory}
\subsection{Density functional theory}
The ground-state electronic energy of an $N$-electron system in a local external potential $v_\text{ext}(\mathbf r)$ is a unique functional of the electron density \cite{hohenberg:1964}. In terms of spin densities ($n^\sigma$, $\sigma=\alpha, \beta$, $n=\sum_\sigma n^\sigma$) the total energy is computed as
\begin{equation}
	E[n^{\alpha}, n^{\beta}]=T_s[n^{\alpha}, n^{\beta}] + \int v_\text{ext}(\bbr) n(\boldsymbol{r}) d\bbr+ J[n] + E_\text{xc}[n^{\alpha}, n^{\beta}]
    \label{eq:Eelec}
\end{equation}
where $T_s$ is the kinetic energy of the non-interacting system, $J$ is the classical Coulomb interaction, and  $E_\text{xc}$ is the XC energy. The ground-state energy is obtained by minimizing the energy over all possible $N$-representable densities.
In KS-DFT, $T_s$ can be exactly computed from the orbitals of the non-interacting system rather than approximated from the density, while $E_\text{xc}$ is always approximated,
\begin{equation}
	T_s[n^{\alpha}, n^{\beta}] = -\frac{1}{2}\sum_{i\sigma} n_i^{\sigma} \langle \psi^{\sigma}_i | \nabla^2 | \psi^{\sigma}_i \rangle
    \label{eqn:KS_TS}
\end{equation}
where $\psi_i^{\sigma}$ are the KS orbitals and $n_i$ the occupation numbers ($\sum_i n_i = N$). While computing $T_s$ directly from the density has a much lower computational cost, existing approximations are too rudimentary to be used for most chemical systems.

In OF-DFT each spin component of the density is expressed in terms of a single bosonic orbital---OF orbital---occupied by $N^{\sigma}$ electrons, and the KEF has to be approximated, typically as the sum of the bosonic kinetic energy---the von Weizsäcker functional---and the Pauli energy which accounts for the fermionic effects:
\begin{eqnarray}
    T_s[n^{\alpha}, n^{\beta}] &=& T_W[n^{\alpha}, n^{\beta}] + T_p[n^{\alpha}, n^{\beta}] \nonumber \\
&=& \frac{1}{8}\sum_{\sigma}\int \frac{|\nabla n^{\sigma}(\bbr)|^2}{n^{\sigma}(\bbr)} + T_p[n^{\alpha}, n^{\beta}]
\end{eqnarray}
Note that the von Weizsäcker functional and Eq.~\eqref{eqn:KS_TS} become the same when evaluated for systems with up to one electron per spin channel, and the exact Pauli functional evaluates to zero in such case.
The minimization of the electronic energy over all $N$-representable densities can be performed with the Lagrange-multipliers method,
\begin{eqnarray}
	L[n^{\alpha}, n^{\beta}] &=& E[n^{\alpha}, n^{\beta}] - \sum_{\sigma}\mu^{\sigma} \left(\int n^{\sigma}(\bbr) d\bbr -N^\sigma\right)  \nonumber \\
 &=& E[n^{\alpha}, n^{\beta}] - \sum_{i\sigma} \varepsilon_{i}^{\sigma} \left(\langle\psi_i^{\sigma}|\psi_i^{\sigma}\rangle -1\right)
\end{eqnarray}
where $\mu$ is the so-called chemical potential and $\varepsilon_{i}^{\sigma}$ are the orbital energies. The functional derivative of $L$ with respect to the orbital $\psi_j^{\sigma}$ is:
\begin{equation}
\begin{aligned}
    |g_j^{\sigma}\rangle &= \frac{\delta L}{\delta |\psi_j^{\sigma}\rangle} = 2 \left[ -\frac{1}{2}\nabla^2 + v_\text{eff}^{\sigma} - \varepsilon_{j}^{\sigma} \right]|\psi_j^{\sigma}\rangle
    \label{eq:Lgradient}
\end{aligned}
\end{equation}
The minimum condition requires the gradient of $L$ to be zero, yielding the regular KS equations:
\begin{gather}
	\frac{\delta L}{ \delta|\psi_j^{\sigma}\rangle} = 0
 \ \ \Rightarrow\ \ 
 \left[-\frac{1}{2}\nabla^2 + v_\text{eff}^{\sigma}\right]|\psi_j^{\sigma}\rangle = \varepsilon_j^{\sigma}|\psi_j^{\sigma}\rangle\\
    v_\text{eff}^\sigma[n^\alpha,n^\beta]=\begin{cases}
		v_\text{ext} + v_H + v_\text{xc}^{\sigma}  & \mbox{KS-DFT} \\
		  v_\text{ext} + v_H + v_\text{xc}^{\sigma} + v_p^{\sigma} & \mbox{OF-DFT}
\end{cases}
\end{gather}
with $v_H[n]$, $v_{xc}[n^\alpha,n^\beta]$ and $v_p[n^\alpha,n^\beta]$ the Hartree, XC and Pauli potentials, the latter only present in OF-DFT. The solution to this set of equations is usually found in a self-consistent-field (SCF) fashion or by direct minimization \cite{helgaker:2000, larsen:2001, jiang:2004}, especially in the OF case.

The functional derivatives of the energy functionals with respect to the density can be easily computed through automatic differentiation~\cite{li:2021}, currently implemented in every major deep-learning package. Throughout the work, XC and Pauli potentials will be computed through the automatic differentiation provided in PyTorch~\cite{NEURIPS2019_9015}
.

\subsection{Learning density functionals}
In the past decade, several works have been devoted to the task of machine-learning density functionals \cite{snyder:2012, snyder:2013, seino:2018, dick:2020,li:2021, ghasemi:2021, kasim:2021, kalita:2021, imoto:2021, kirkpatrick:2021,wu:2022}, especially the XC functional. It became clear early on \cite{snyder:2012} that training a density functional by just matching the energies for the reference densities produced functionals which are unusable in actual SCF calculations. The reason for this is that when training only by energy, the functional has little chance of learning the potential, i.e. the functional derivative, so the minimum condition expressed in equation Eq.~\eqref{eq:Lgradient} is not necessarily fulfilled for the reference ground-state density. 

Li. \textit{et al.} \cite{li:2021} proposed to use the KS equations themselves as regularizers, matching the converged energies and densities from an SCF calculation to the reference ones. The regularization comes from the fact that the backward pass traverses the entire SCF calculation and that the losses are computed from converged quantities. The main drawback of the method is that it requires a full SCF calculation for each sample of the training set every epoch, which becomes extremely expensive for the large datasets that are needed for realistic functional training. 

Kirkpatrick \textit{et al.} \cite{kirkpatrick:2021} avoided this problem for training their DM21 XC functional by minimizing the second-order change of the energy from a single SCF step from the reference density. This forces the functional to have a stationary point close to the reference density. While this method has been proven to be highly efficient, it cannot be applied in the OF case because it requires the knowledge of reference virtual orbitals, which are unavailable for the OF Hamiltonian. Nonetheless, the essential implication is that the regularization merely has to induce the reference densities to be the energy-functional minima.

Building upon this concept, we propose utilizing directly the gradient norm in Eq.~\eqref{eq:Lgradient} as a regularizer in our work. In a conventional DFT calculation, this gradient directs the search for the orbitals (density) that minimize the electronic energy. When training a functional, this gradient will instead steer the search for the Hamiltonian whose eigenfunctions are the reference orbitals (density), by means of optimizing the Pauli or XC functional. Consequently, the overall loss function will be a combination of an energy and gradient regularization terms:
\begin{equation}
    \mathcal{L} = \sqrt{\mathbb{E}[(E - E^*)^2 / N] + \lambda \mathbb{E}[\langle g|g\rangle]}
    \label{eq:loss_fn}
\end{equation}
where $E$ is the total energy, computed from Eq.~\eqref{eq:Eelec} using the ML functional, $E^*$ is the reference total energy, and $|g\rangle$ is the gradient from Eq.~\eqref{eq:Lgradient} with the effective potential, $v_\text{eff}^{\sigma}$, computed on the reference orbitals. The orbital energy ($\epsilon_i^{\sigma}$) therein is just the expectation value of the Fock matrix calculated with the ML functional. Finally, $\lambda$ controls the effect of the regularization in the training process. This regularization is applicable to train the KEF or XC functional both in OF-DFT and KS-DFT. The regularization term is computed, for every sample, as the sum of the norm of the gradients on each orbital,
\begin{eqnarray}
	\langle g |g\rangle &=& \sum_{i\sigma} \langle g_{i}^{\sigma} |g_{i}^{\sigma}\rangle
\end{eqnarray}
These gradients can be projected onto the basis-set elements in which $\psi_i$ is expressed ($\{\phi_a\}$):
\begin{eqnarray}
    \psi_i(\mathbf r) &=&\sum_j c_{ij} \phi_j(\mathbf r)\\
    \langle \phi_a | g_i^{\sigma} \rangle &=& \sum_j c_{ij} \left[ F_{aj}^{\sigma} - \varepsilon_i^{\sigma} S_{aj}\right]
\end{eqnarray}
where $\mathbf{F}$ and $\mathbf{S}$ are the Fock and overlap matrices, respectively.

\section{\label{sec:results} Results}

We demonstrate the effectiveness of the new regularization approach on a set of representative systems, for which we train the KEF. These systems include linear hydrogen chains in 1D, non-interacting electrons in 1D Gaussian external potentials, and free atoms in 3D. Furthermore, on free atoms, we also compare learning the Pauli and XC functionals.

In order to learn the Pauli functional, we create a training set consisting of target energies and OF orbitals that generate the desired density. To obtain the target energies and densities, we utilize KS-DFT, which enables us to compute the exact kinetic energy of the non-interacting system. The reference OF orbitals are determined by projecting the square root of the density onto the basis-set elements and solving a set of linear equations given by
\begin{equation}
    \langle \phi_a | \sqrt{n} \rangle = \sum_i c_i \langle \phi_a | \phi_i \rangle = \sum_i c_i S_{ai}
    \label{eq:OFMOcoef}
\end{equation}

To learn the XC functional, we use reference densities obtained from another functional. According to Kirkpatrick \textit{et al.} \cite{kirkpatrick:2021}, the energy computed from a density functional is relatively insensitive to minor changes in the densities, that is, the density error on the tested functionals is low. Hence, densities converged with different functionals yield similar electron energies. On the other hand, energy driven error appears to be quite large, as the energy calculated from a single density varies significantly depending on the density functional. For more details, we refer the reader to their work.

The ML functionals, whether KEF or XC functional, are constructed as integrals over the entire space of an energy density,
\begin{equation}
    F[n^{\alpha}, n^{\beta}] = \int f_{\boldsymbol\theta}\big(\mathbf z[n^{\alpha}, n^{\beta}](\mathbf r)\big) n(\mathbf r) d\mathbf r
\end{equation}
A multilayer perceptron (MLP), $f_{\boldsymbol\theta}$, is used to represent the energy density, and is fed with features $\mathbf z$ computed from the electron density. In all cases the input features are heavily inspired by those from Li. \textit{et al.}~\cite{li:2021}, and are detailed for each individual system type below.

\subsection{1D hydrogen chains}
The 1D hydrogen chain consists on a set of equidistant hydrogen atoms aligned along one axis. The external potential is modeled by the exponential Coulomb interaction \cite{baker:2015}:
\begin{equation}
    v_\text{ext}(x) = -Z v_\text{exp}(x), \quad v_\text{exp}(x) = A \exp(-\kappa|x|)
	\label{eq:softcoulomb}
\end{equation}
with an unpolarized LDA exchange from Ref.~\cite{elliot:2014} and correlation term from Ref.~\cite{baker:2015}. The classical Coulomb interaction takes the form:
\begin{eqnarray}
    J[n] &=& \frac{1}{2}\iint n(x) n(x') v_\text{exp}(x-x') dxdx'\\
    v_H(x) = \frac{\delta J}{\delta n} &=& \int n(x') v_\text{exp}(x-x') dx'
\end{eqnarray}

The features $\mathbf z$ fed to the MLP representing the non-local Pauli functional are:
\begin{eqnarray}
	z_1 &=& \log(n(x) + 10^{-4})\\
	z_i &=& \log(G[n](x;\alpha_i) + 10^{-4}),\quad i > 1
\end{eqnarray}
where $G$ is a global convolution of the density with a Gaussian kernel:
\begin{equation}
    G[n](x; \alpha) = \frac{1}{\sqrt{2 \pi \alpha}} \int_{-\infty}^{\infty} n(x') \exp\left(-\frac{(x-x')^2}{2 \alpha}\right) dx'
	\label{eq:globalconv_grid}
\end{equation}
The degree of non-locality of the kernel width is governed by the $\alpha$ parameter. Larger values of $\alpha$ result in more non-local features.

The training set consists of KS-DFT densities and energies computed with the XC terms referenced above, for equidistant H$_n$ chains ($n=2,\, 4,\, 6,\, 8$). The interatomic distances included in the training set are reported in the SI. In the following sections, we investigate how the performance of the trained Pauli functional is influenced by the number of training data, the number of convolutions, and the inclusion of non-local features. In all cases we compare with our reference KS-DFT calculations.

\begin{figure*}[hbtp]
\includegraphics[width=\linewidth]{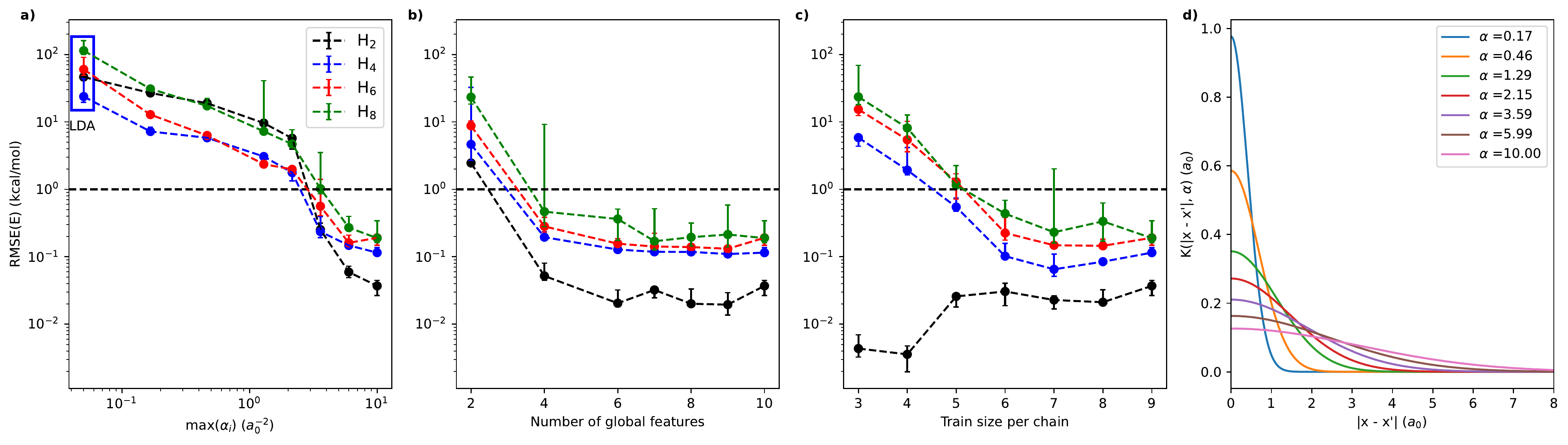}
\caption{\label{fig:1} Validation errors for the hydrogen chains with respect to (a) the maximum width of the global convolution, (b) the number of global features fed to the MLP, and (c) the number of training samples per hydrogen chain. The dots represent the median error of ten models and the error bars the Q1 and Q3 errors, in a box-plot fashion. Chemical accuracy is shown with a dashed horizontal line. Pauli LDA results in (a) are obtained for a model with no global features. (d) Sample of Gaussian kernels in Eq.~\eqref{eq:globalconv_grid} for different $\alpha$ values.}
\end{figure*}

\subsubsection{Number of convolutions and degree of non-locality}
\begin{figure}
\includegraphics[width=\linewidth]{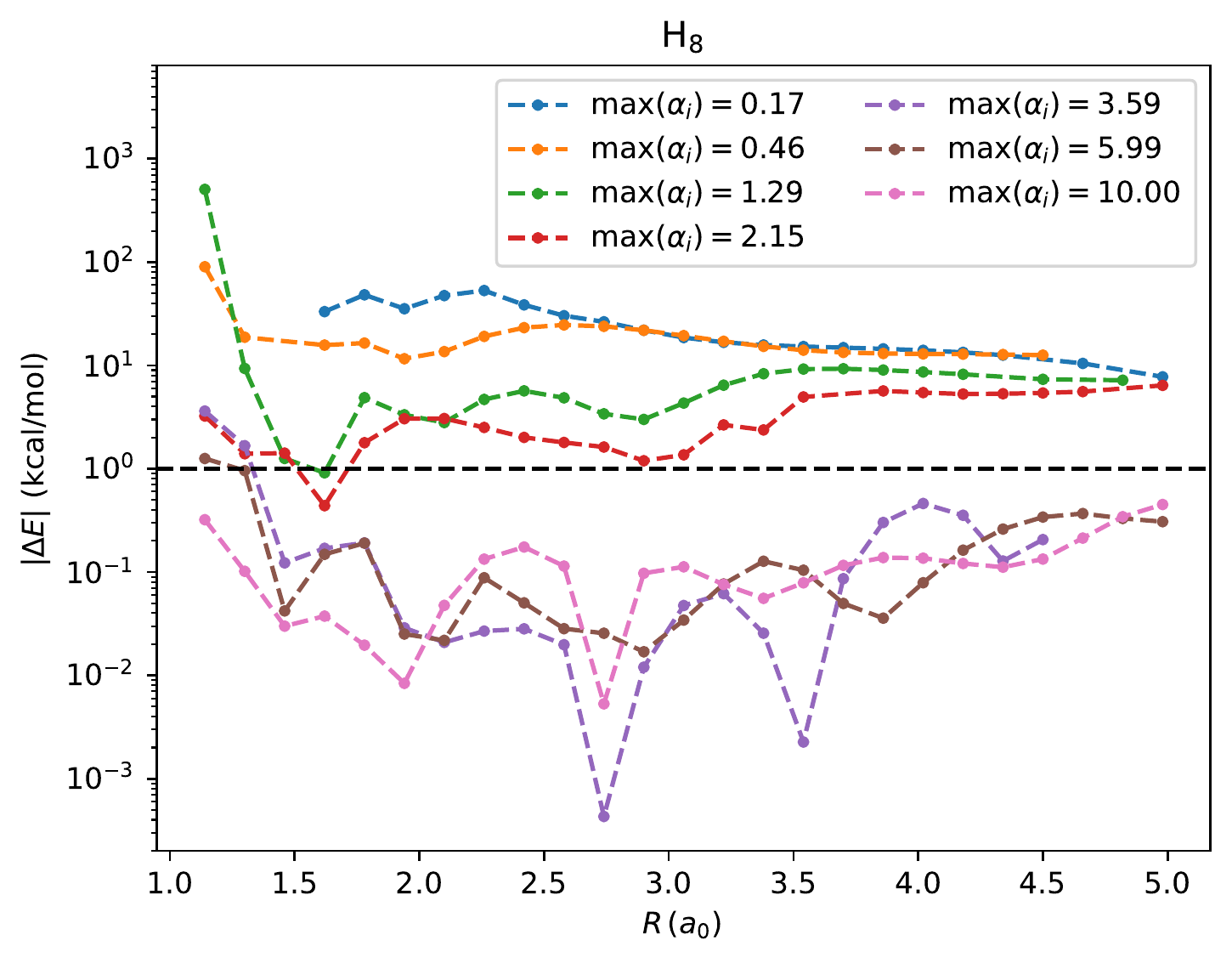}
\caption{\label{fig:2} Energy errors along the H$_8$ potential energy curve for models with increasing non-local features. $R$ indicates the interatomic distance between H atoms.}
\end{figure}

The non-interacting kinetic energy is known for its highly non-local nature \cite{garcia:1996}. While even the LDA approximation of the XC functional is able to produce accurate densities, semi-local approximations of the kinetic energy often struggle to reproduce accurate energies and densities. In our model, we account for non-local effects in the Pauli functional through global convolutions that compute non-local features. We aim to determine the minimum amount of non-locality required by the functional to achieve accuracy by training different models with increasing numbers of global features and larger kernel widths, which introduce more non-local information into the model. The relationship between non-locality of features and accuracy is evident in Fig.~\ref{fig:1}a. Further insight into the impact of non-local effects can be obtained by examining the errors in the potential energy curve of H$_8$, as shown in Fig.~\ref{fig:2}. When $\max(\alpha) < 3 a_0^{-2}$, errors below chemical accuracy are rare. For larger convolutional kernels, accuracy improves significantly in the dissociation region, but larger errors are still present for smaller $R$ values. Non-locality has a more pronounced effect at lower $R$ values, with more compressed and correlated densities, and the remaining accuracy is achieved through wider global convolutions.

The increase in accuracy can be attributed to the inclusion of non-locality in the global features, as depicted in Fig.~\ref{fig:1}b. A series of global functionals are trained with an equal $\alpha$ range $\alpha \in [0.1, 10.0]$, while increasing the number of non-local features. The results indicate that the functional can attain chemical accuracy on all test systems after four or six global features. This finding suggests that the improved accuracy is mainly due to the range of non-local features rather than their number.

Note that a simple LDA approximation for the KEF is not flexible enough to be simultaneously zero for H$_2$, where the vW is exact, and different from zero for the chains with more than two electrons.

\subsubsection{Amount of training data}
Ten models were trained on an increasing amount of data, from one geometry per chain to nine of them, to determine how the training-set size affects the validation performance. The effect of the training-set size on the validation error for several hydrogen chains is presented in Fig.~\ref{fig:1}c. In all cases, the models reached chemical accuracy for six training points, after which the error stabilised. This behaviour is mainly due to the fact that the model no longer has to extrapolate out of the range of $R$ values seen in the training set (the model is still validated on unseen values of $R$). This demonstrates a difference between the XC functional and non-interacting KEF. Li \textit{et al.} \cite{li:2021} showed in their work that about two training points for each H$_2$ and H$_4$ are needed to accurately recover the potential energy curve, demonstrating great generalisation power. In contrast, training the non-interacting KEF is expected to require much more data. 

\begin{figure*}[hbpt]
\includegraphics[width=\linewidth]{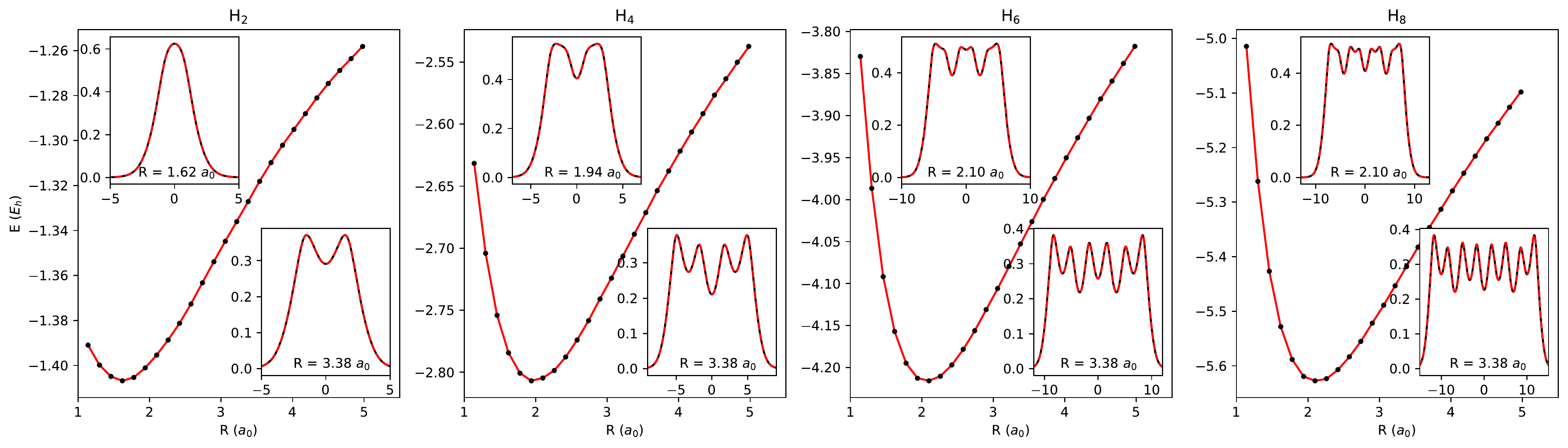}
\caption{\label{fig:3} Potential energy curves of the hydrogen chains. Reference KS-DFT energies (black dots) and converged OF-DFT energies (red lines) are shown. The insets show the KS-DFT density in black compared to the converged OF-DFT density. The top insets show the density close to the minimum in the potential energy curve and bottom one the density for a geometry closer to dissociation. The $x$-axis of the inset represents the $x$-coordinate in atomic units and the $y$-axis the electron density ($n(x)$).}
\end{figure*}

The potential energy curves of the hydrogen chains are presented in Fig.~\ref{fig:3} together with some selected densities, close to their respective minima and tending to dissociation. Both energies and densities have been obtained via an SCF calculation employing the trained functional.

\subsection{Non-interacting electrons}
The next system is the one used by Snyder \textit{et al.} \cite{snyder:2012} in their original work, which consists of a set of $N$ non-interacting, spinless electrons confined in a box, $x \in [0,1]$, feeling an external potential expressed as the sum of three Gaussian functions, generated by randomly sampling $a \in [1, 10]$, $b \in [0.4, 0.6]$  and $c \in [0.03,0.10]$:
\begin{eqnarray}
    v_\text{ext}(x) = \sum_{i=1}^3 a \exp \left(-\frac{(x - b)^2}{2 c^2} \right)
\end{eqnarray}
 The Pauli functional is trained for different number of electrons, $N=2,\,3,\,4$, on a set of 100 random potentials. A final model is trained on 400 random potentials with $N \in [1,4]$. These models are validated on a different set of 1000 external potentials. The features fed to the MLP are:
\begin{eqnarray}
	z_1 &=& n(x)\\
	z_i &=& G[n](x;\alpha_i),\quad i > 1
\end{eqnarray}
with the global convolution in Eq.~\eqref{eq:globalconv_grid}. The validation energy and density errors are presented in Table~\ref{tab:table1}. The density error is computed as:
\begin{equation}
    (\Delta n)^2 = \int_{0}^1 \left(n(x) - n'(x) \right)^2 dx
\end{equation}
where $n(x)$ and $n'(x)$ are the reference KS-DFT and OF-DFT densities, respectively.

\begin{figure*}[hbpt]
\includegraphics[width=\linewidth]{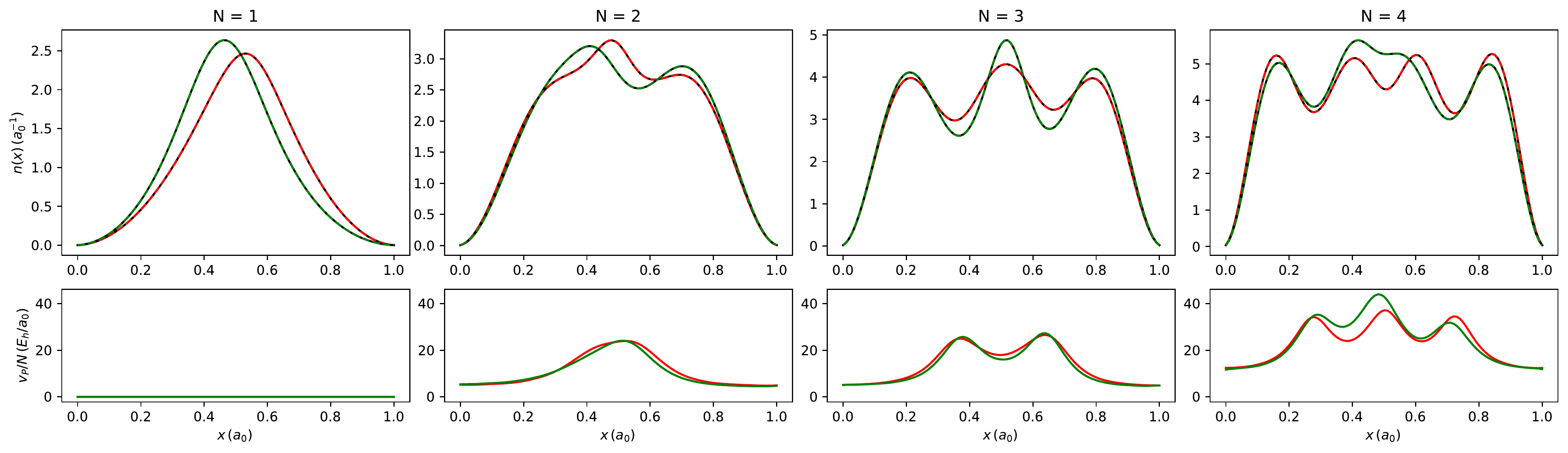}
\caption{\label{fig:4} Top panels: samples of KS densities (black solid lines) and the corresponding converged OF densities (red and green dashed lines) for two randomly selected validation external potentials. Bottom panels: Pauli potentials for the converged densities.}
\end{figure*}

\begin{table}[hbpt]
\caption{\label{tab:table1} Validation errors on a set of 1000 randomly generated external potentials for converged OF-DFT calculations with the trained Pauli functional. Energy errors in kcal/mol and density errors in $a_0^{-1}$.}
\begin{ruledtabular}
\begin{tabular}{c c c c c c c}
$N$ & $\overline{|\Delta E|}$& $|\Delta E|^{std}$ & $|\Delta E|^{max}$ & $\overline{\Delta n}$ & $\Delta n^{std}$ & $\Delta n^{max}$\\
\hline
2 & 0.10 & 0.15 & 2.51 & $3.65 \times 10^{-7}$ & $2.33 \time 10 ^{-6}$ & $6.51 \times 10^{-5}$\\
3 & 0.26 & 0.39 & 6.86 & $2.00 \times 10^{-6}$ & $3.00 \time 10 ^{-6}$ & $3.70 \times 10^{-5}$ \\
4 & 0.48 & 0.72 & 8.10 & $4.33 \times 10^{-6}$ & $8.60 \time 10 ^{-6}$ & $1.01 \times 10^{-4}$\\
1---4 & 0.43 & 0.78 & 12.05 & $3.71 \times 10^{-6}$ & $1.72 \time 10 ^{-5}$ & $3.17 \times 10^{-4}$\\
\end{tabular}
\end{ruledtabular}
\end{table}
Note that the errors in Table~\ref{tab:table1} are computed with the converged densities and energies from a regular SCF calculation since the Pauli potential becomes a smooth function of the density thanks to the gradient regularization. Fig.~\ref{fig:4} shows the converged densities for several randomly selected external potentials with the number of electrons varying from 1 to 4. The Pauli potentials, computed from the converged OF densities, are presented in the bottom panels, showing a smooth behavior along the $x$ coordinate.

\subsection{Atoms}
Finally, we validate our methodology for learning the KEF and XC functional on the atoms from the first two periods of the periodic table, extending the work from Ghasemi and Kühne~\cite{ghasemi:2021} to a larger set of atoms and to the spin-polarized DFT framework. Atoms can be treated as an effective one-dimensional problem since only the radial part of the density has to be solved:
\begin{equation}
    n(\mathbf r) = n(r) n(\theta, \varphi), \quad n(\theta, \varphi) = 1
\end{equation}
Since we are considering multiple spin states, we will utilize the spin-polarized DFT in this scenario. The external and Coulomb potentials\cite{theoreticalphysics} are computed using the regular Coulomb interactions between charged particles:
\begin{eqnarray}
	v_{ext}(r) &=& - \frac{Z}{r}\\
	v_H(r) &=&  4 \pi \int_0 ^{\infty} \frac{n(r')r'^2}{|r-r'|} dr' \nonumber\\
        &=& 4 \pi \left[ \frac{1}{r}\int_0^r n(r') r'^2 dr' + 
    \int_r^\infty n(r')r'dr'\right]
\end{eqnarray}
Pauli and XC energies are computed as:
\begin{equation}
	E_F[n^{\alpha}, n^{\beta}] = 4 \pi \int_0^\infty f[n^{\alpha}, n^{\beta}](r) n(r) r^2 dr
\end{equation}
where $f[n^{\alpha}, n^{\beta}]$ is the Pauli or XC energy density.

For a spherically symmetric external potential the Schrödinger equation can be written as a one-dimensional equation,
\begin{equation}
	\left[-\dfrac{1}{2}\dfrac{d^2}{dr^2} + \frac{l(l+1)}{2r^2} + v^{\sigma}(r) \right] u^{\sigma}(r) = \varepsilon^{\sigma} u^{\sigma}(r)
\end{equation}
with $\psi^{\sigma}(r) = u^{\sigma}(r) / r$, $l$ the orbital angular momentum, and $v^{\sigma}(r) = v_H(r) + v^{\sigma}_{xc}(r) + [v^{\sigma}_P(r)]$, the latter potential term only present in OF-DFT. In OF-DFT case no centrifugal term is considered, that is $l=0$, since this is a kinetic-energy contribution which is described by the Pauli functional.
The features fed to the MLP representing the functional energy density are:
\begin{eqnarray}
	z_1 &=& \log(n^{\alpha}(r) + 10^{-4})\\
	z_2 &=& \log(n^{\beta}(r) + 10^{-4})\\
	z_{i_\alpha} &=& G[n^\alpha](r; \alpha_{i_\alpha})\\
	z_{i_\beta} &=& G[n^\beta](r;\alpha_{i_\beta})
\end{eqnarray}
where the convolution widths $\alpha_i$ are the same for $\alpha$ and $\beta$ densities and the convolution is computed as
\begin{equation}
    G[n](r;\alpha) = \frac{1}{\sqrt{2 \pi \alpha}} \int_{0}^{\infty} n(r') \exp\left(-\frac{(r-r')^2}{2 \alpha}\right) dr'
\end{equation}
We demand that the XC energy density is always negative, so the output of the MLP is multiplied by $-1$ in this case. In order to make the energy density spin-symmetric, the same MLP is run twice and averaged for both spin orderings.

Similar to the previous test conducted on hydrogen chains, we evaluate the accuracy of Pauli and XC functionals with respect to the level of non-locality incorporated in the global features. In both cases the training set will consist of energies and densities of the elements in the first two periods of the periodic table. The reference energies and densities for the Pauli functional are computed with UKS/LDA. To learn the XC functional, the reference energies are computed with UCCSD(T)/cc-pVTZ using PySCF\cite{pyscf:2018, pyscf:2020}. UKS/LDA reference densities are taken in this case, similar to how DM21 was trained, leveraging the fact that the XC functionals are, in general, little sensitive to the reference densities\cite{kirkpatrick:2021}.

\begin{figure}[hbpt]
\includegraphics[width=0.9\linewidth]{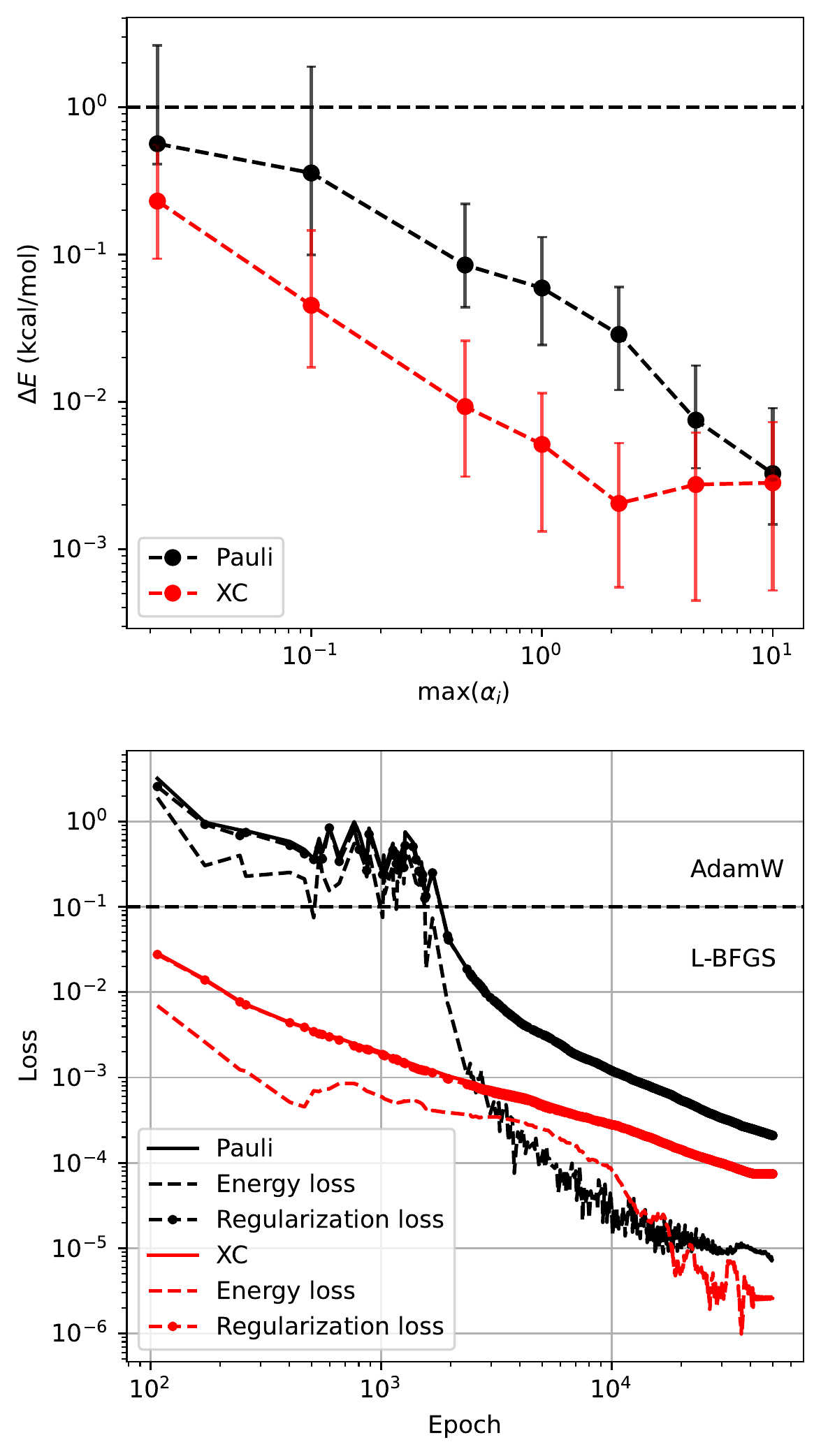}
\caption{\label{fig:5} Comparison of the median errors of 10 training runs of the Pauli and XC functionals for the atoms in the first two periods of the periodic table against the amount of non-locality in the global features. The black points represent the average values of the median for all the atoms. Training loss curves for the largest model, together with the energy and regularization contributions, for the Pauli functional (black) and the XC functional (red).}
\end{figure}

The accuracy achieved by the Pauli and XC functionals with respect to the level of non-locality incorporated in the input features is compared in Fig.~\ref{fig:5}. While the XC functional needs little non-locality, reaching an excellent accuracy with only two narrow global convolutions, the Pauli functional does not produce accurate results until large convolutional widths are employed to construct the density features. The learning curves presented in the right panel of Fig.~\ref{fig:5} already show a difference between both functional learning, where the Pauli functional needs a significantly larger amount of epochs to reach similar loss values than the XC functional. The regularization term is learned significantly faster for the XC functional, suggesting a less intricate structure of the underlying true functional, compared to the KEF.

The small amount of training data raises the question of how much of this precision is actually physics learned by the functional or pure overfitting. To answer this question we compute the ionization energies of the atoms with our learned functionals, since the cationic species were never seen in the training process. In Table~\ref{tab:ionization} the ionization energies are presented for each atom compared to the reference energy in each case, \textit{i.e.} KS-DFT/LDA energies to validate the Pauli functional and UCCSD(T) energies to validate the XC functional. These results present some clear overfitting in the case of Pauli functional, where in most cases it is not even possible to converge the SCF calculation of the cation, and those that converge present large energy differences. Although the learned XC functional is much more robust with respect to convergence, it also exhibits deviations from the actual ionization energies.

\begin{table*}[hbpt]
\caption{\label{tab:ionization} Ionization energies of the atomic species in kcal/mol. For the learned functionals the energy difference (in kcal/mol) with respect to the reference ionization energy is presented.}
\begin{ruledtabular}
\begin{tabular}{c | c r r r r r || c r }
Atom & KS-DFT/LDA-PW92 & OF-DFT/Trained & $\tau_{max} = 0.0$ & $\tau_{max} = 0.5$ & $\tau_{max} = 1.0$ & $\tau_{max} = 1.5$ & CCSD(T)/cc-pVTZ & KS-DFT/Trained \\\hline
He & 560.16 & $ -0.02 $&   $   -0.11$ &$   0.01$& $ -0.01$& $ -0.04$ & 565.58  & $+ 6.21$ \\
Li & 125.40 & $ 80.59 $&   $   84.69$ &$  86.64$& $  6.48$& $  0.76$ & 123.44  & $+25.24$ \\
Be & 208.11 & $-      $&   $-1382.45$ &$  116.92$   & $ -0.86$& $ -2.96$ & 214.11  & $+25.61$ \\
B  & 197.71 & $+106.53$ &  $  122.69$ &$  10.11$& $ -0.67$& $  0.55$ & 189.74  & $+41.73$ \\
C  & 271.25 & $+141.81$ &  $  166.88$ &$  61.14$& $  5.98$& $ -0.22$ & 257.92  & $+50.29$ \\
N  & 345.78 & $+186.73$ &  $  239.37$ &$  55.26$& $  -$& $ -0.95$ & 333.21  & $+52.79$ \\
O  & 320.51 & $-      $ &  $  185.46$ &$ 102.35$& $  3.43$& $  4.27$ & 307.29  & $+50.58$ \\
F  & 416.47 & $+416.70$ &  $   -    $ &$ 193.64$& $   -1.80$  & $ -3.84$ & 395.30  & $+58.68$ \\
Ne & 511.53 & $+ 475.4$8 & $   -    $ &$ 434.67$& $  9.52$& $ -1.57$ & 491.41  & $+60.11$\\
\end{tabular}
\end{ruledtabular}
\end{table*}

To test how the training of the Pauli functional would proceed in a more realistic scenario with a larger amount of training data, we generate larger datasets of atoms with fractional nuclear charges, under the external potential
\begin{equation}
    v_\text{ext}(r) = - \frac{Z + \tau}{r}
\end{equation}
where $\tau$ is a random number between $-0.1$ and $\tau_\text{max}$. Four datasets are generated with $\tau_\text{max} = 0.0,\, 0.5,\, 1.0$, and $1.5$. The datasets contain a total number of 400 atoms with artificial nuclear charges, 40 samples per $Z$. The spin state is that for the ground of state of the atom with atomic number $Z$. In Table~\ref{tab:ionization} the energy errors in the ionization energy are presented for each atom. The two datasets with larger $\tau_\text{max}$ provide much more information for the functional to gain physical insight about the true KEF. Note that these datasets do not necessarily contain the neutral and cation species, yet the functional becomes general enough as to accurately predict them. 

\section{\label{sec:conclusions} Conclusions}
In this work we present a new method to regularize the training of density functionals, derived from the variational condition that the global energy functional presents a minima for the ground electronic state density. In particular, the gradient of the Lagrangian with respect to the density is employed to guide the search of the functional which fulfills the minimum condition, making them applicable to regular SCF calculations. This methodology is directly applicable to the training of KEF and XC functional. This is a cheap regularization technique compared to performing a full SCF calculation for each of the training samples,  and we anticipate that it can have widespread applicability in realistic functional training scenarios.

It is evident that forthcoming attempts to train the KEF will heavily depend on the inclusion of non-local features, which we have found to be more crucial than the number of global features and even the size of the network. The KEF reveals a more complex inner structure to learn, compared to the XC functional, as can be extracted from the regularization learning curves for both functionals with equivalent ML models and training schemes.
Additionally, our findings reveal that the KEF is highly susceptible to overfitting on small datasets, as evidenced by its inability to accurately replicate the ionization energy of atoms. In contrast, the XC functional, which was solely trained on neutral atoms, displays remarkable robustness. Notably, all the SCF calculations converge to reasonable energies for the XC, while the KEF exhibits a dismal performance. Another instance where the KEF displays less generality than the XC is demonstrated in the hydrogen chain, where a greater number of training points was required for the KEF to accurately recover the potential energy curve compared to the XC case presented by Li \textit{et al.} Therefore, it is anticipated that the datasets used to train the KEF must be substantially larger to enable the acquisition of sufficient physical knowledge. Alternatively, additional regularization techniques such as equipping the ML functional with exact scaling properties could be also used.


%
%

%

\bibliography{bibliography}

\begin{acknowledgments}
We thank P.\ Gori-Giorgi for valuable comments on the manuscript.
Funding is acknowledged from the German Ministry for Education and Research (Berlin Institute for the Foundations of Learning and Data, BIFOLD). P.M.S. acknowledges funding from MICINN (Spain) under grants PID2021-122549NB-C21 and -C22.
\end{acknowledgments}

\end{document}


\maketitle
\tableofcontents

\section{Hydrogen chains}
\subsection{Training set}

The training set consists on a set KS energies and densities for several equidistant hydrogen chains, separated by a distance $R$. The values of $R$ selected as training set are:

\begin{table}[hbtp]
	\centering
	\caption{\label{tab:training_set} Interatomic distances for each number of geometries per hydrogen chain.}
	\begin{tabular}{c | l}
		Geometries per chain & $R\, (a_0)$\\\hline
		1 & 2.0\\
		2 & 2.0, 3.0\\
		3 & 1.0, 2.0, 3.0\\
		4 & 1.0, 2.0, 3.0, 4.0\\
		5 & 1.0, 2.0, 3.0, 4.0, 5.0\\
		6 & 1.0, 2.0, 2.5, 3.0, 4.0, 5.0\\
		7 & 1.0, 2.0, 2.5, 3.0, 3.5, 4.0, 5.0\\
		8 & 1.0, 2.0, 2.5, 3.0, 3.5, 4.0, 4.5, 5.0\\
		9 & 1.0, 1.5, 2.0, 2.5, 3.0, 3.5, 4.0, 4.5, 5.0\\
	\end{tabular}
\end{table}

The densities are expressed in an uniform grid in the range $[-30,30]\, a_0$ with $dx=0.1$.

\subsection{General training procedure}

First the number of training points per hydrogen chain is selected, from 1 to 9, as presented in table~\ref{tab:training_set}.

The Pauli functional is initialized, specifying the number of global features to be used and the range of $\alpha$ values. In our case, $\alpha$ parameters are not learnable parameters and are initialized as equidistant values in logarithm space between two specified limits, which in this experiments are, $\min(\alpha) = 0.1$ and $\max(\alpha) = 10.0$.

The MLP has a fixed structure across all our experiments, with two hidden layers and SiLU non linearities, except for the output layer, which uses Softplus, to guarantee the positiveness of the energy density. Every hidden layer has a dimension of 60. The input layer has a size of $N_{glob} + 1$, where $N_{glob}$ is the number of non local features and the extra dimension is the logarithm of the density, as discussed in the main text.

An L-BFGS optimizer is employed in the training process with line search. A maximum number of 20k epochs is used.

The regularization weight $\lambda$ is always taken to be one.

\subsection{Effect of non local features}

In this experiment we test what is the effect of increasing the amount of non locality in the global features fed to the MLP representing the functional energy density. For that, we increasingly add more features generated from wider convolutions.\\

The full set of $\alpha$ parameters in the gaussian kernels is computed as equidistant points in logarithm space between $10^{-1}$ and $10$:
\begin{verbatim}
	alpha = torch.logspace(-1, 1, 10)
\end{verbatim}
The $\alpha$ parameters used in this set of experiments are:
\begin{itemize}
	\item $\max(\alpha_i) = 0.17$: $[0.10, 0.17]$
	\item $\max(\alpha_i) = 0.46$: $[0.10, 0.17, 0.28, 0.46]$
	\item $\max(\alpha_i) = 1.29$: $[0.10, 0.17, 0.28, 0.46, 0.77, 1.29]$
	\item $\max(\alpha_i) = 2.15$: $[0.10, 0.17, 0.28, 0.46, 0.77, 1.29, 2.15]$
	\item $\max(\alpha_i) = 3.59$: $[0.10, 0.17, 0.28, 0.46, 0.77, 1.29, 2.15, 3.59]$
	\item $\max(\alpha_i) = 5.99$: $[0.10, 0.17, 0.28, 0.46, 0.77, 1.29, 2.15, 3.59, 5.99]$
	\item $\max(\alpha_i) = 10.00$: $[0.10, 0.17, 0.28, 0.46, 0.77, 1.29, 2.15, 3.59, 5.99, 10.00]$
\end{itemize}

The training set consists on nine geometries for each hydrogen chain.

For each set of $\alpha$ parameters we run 10 training processes according to the setup specified in the General training procedure section.

\subsection{Effect of the number of convolutions}

Contrary to the above experiment where the range of the widths in the gaussian kernel is not constant, in this experiment $\alpha \in [10^{-1}, 10]$ always:
\begin{table}[hbtp]
	\centering
	\begin{tabular}{c | l}
		Number of non local features & $\alpha$ \\\hline
		2  & 0.10, 10.00\\
		4  & 0.10,  0.46,  2.15, 10.00\\
		6  & 0.10,  0.25,  0.63,  1.58, 3.98, 10.00\\
		7  & 0.10,  0.22,  0.46,  1.00, 2.15,  4.64, 10.00\\
		8  & 0.10,  0.19,  0.37,  0.72, 1.39,  2.68,  5.18, 10.00\\
		9  & 0.10,  0.18,  0.32,  0.56, 1.00,  1.78,  3.16,10.00\\
		10 & 0.10, 0.17, 0.28, 0.46, 0.77, 1.29, 2.15, 3.59, 5.99, 10.00\\
	\end{tabular}
\end{table}

In general, the exact $\alpha$ values can be computed as:
\begin{verbatim}
	alpha = torch.logspace(-1, 1, N)
\end{verbatim}
where $N$ in the number of non local features.

For each set of $\alpha$ parameters we run 10 training processes according to the setup specified in the General training procedure section.
\subsection{Effect of the training set}

To measure the effect of the training set size in the accuracy of the Pauli functional 10 models are trained for each number of training points in table~\ref{tab:training_set} with the training setup specified in the General training procedure section.

\section{Non-interacting electrons}
\subsection{General training procedure and validation}
Given a number of electrons a set of 100 reference energies and densities are computed on a gaussian external potential with random parameters $a$, $b$ and $c$.

The Pauli functional is initialized as an MLP with the exact same architecture as the one commented for the hydrogen chain. Only the input features are different, since in this case the density itself is fed to the MLP and not the natural logarithm of it. The same is true for the global features.

The number of global features is fixed to 6 in this system and the $\alpha$ values for the non local features are equidistant in logarithm space between $\min(\alpha) = 10^{-3}$ and $\min(\alpha) = 1$.

For this system we found the most stable training process starting with an AdamW optimizer, with an initial learning rate of $10^{-2}$ and a learning rate scheduler which decreases it by a factor of 10 if the loss is not improved in 300 epochs. Once the loss reaches a value threshold, the optimizer is switched to L-BFGS with line search, since is significantly boosts the training process. This loss threshold is different for each number of electrons

\begin{table}[hbtp]
	\centering
	\begin{tabular}{c | c}
		N & Loss threshold\\\hline
		2 & 0.2 \\
		3 & 0.5 \\
		4 & 0.8 \\
		1---4 & 0.8 \\
	\end{tabular}
\end{table}

The regularization weight $\lambda$ is always taken to be one.

The results reported in the paper correspond to the training of a single model for each of the number of electrons. The validation errors are computed on a set of newly generated 1000 gaussian potential with random parameters.

\section{Atoms}
\subsection{Effect of non local features}
The Pauli functional is trained on the atoms from the first two periods of the periodic table, from H to Ne. The training set of energies and densities is computed from an UKS/PW92-LDA calculation. The OF orbital coefficients are calculated from the KS reference:
\begin{equation}
	C^{\sigma} = \diag(P^{\sigma} / N^{\sigma})
\end{equation}
where $P^{\sigma}$ is the KS density matrix for $\sigma \in {\alpha, \beta}$ spin channel and $N^{\sigma}$ is the number of electrons in the spin channel. In the particular case of the $\beta$ channel in H
\begin{verbatim}
	C[:] = 0.0
\end{verbatim}
The OF density matrix is:
\begin{equation}
	P^{\sigma} = N^{\sigma} C^{\dagger} C
\end{equation}

For the XC functional, the orbital coefficients from the UKS/PW92-LDA calculation are taken as ground truth together with UCCSD(T)/cc-pVTZ energies. Both functional are trained with the same setup.

A multilayer perceptron is initialized with four hidden layers and 60 neurons per layer. For the XC functional the negative transform of the energy density is turned on to guarantee its negativeness. The number of global features is increased from 2 to 10, with the following $\alpha$ values:
\begin{itemize}
	\item $\max(\alpha_i) = 0.02$: $[0.01,  0.02]$
	\item $\max(\alpha_i) = 0.10$: $[0.01,  0.02,  0.05,  0.10]$
	\item $\max(\alpha_i) = 0.46$: $[0.01,  0.02,  0.05,  0.10, 0.22,  0.46]$
	\item $\max(\alpha_i) = 1.00$: $[0.01,  0.02,  0.05,  0.10, 0.22,  0.46,  1.00]$
	\item $\max(\alpha_i) = 2.15$: $[0.01,  0.02,  0.05,  0.10, 0.22,  0.46,  1.00,  2.15]$
	\item $\max(\alpha_i) = 4.64$: $[0.01,  0.02,  0.05,  0.10, 0.22,  0.46,  1.00,  2.15, 4.64]$
	\item $\max(\alpha_i) = 10.00$: $[0.01,  0.02,  0.05,  0.10, 0.22,  0.46,  1.00,  2.15, 4.64, 10.00]$
\end{itemize}
The training process runs for 5k steps, starting with an AdamW optimizer with a learning rate of 0.01 and switching to L-BFGS with line search as the loss becomes lower than 0.1.

\subsection{Ionization energies}
The ionization energies are computed as the energy difference of the neutral and positively charged species:
\begin{equation}
	I(A) = E(A^+) - E(A)
\end{equation}
For the Pauli functional the reference energies will be UKS/PW92-LDA while for the XC it is UCCSD(T)/cc-pVTZ.

To get the dataset of fractional nuclear charges a list of Z values is created from $Z=1$ to $Z=10$. Each $Z$ value is repeated 40 times. Next, a random array $\tau$ with shape 400 is generated with values in $[-0.1, \tau_{max}]$. The fractional nuclear charge is set to the external potential and the KS equations are solved, generating the dataset of fractional nuclear charges.

This training set is used to train a larger model with 4 hidden layers with 120 neurons per layer and 10 non-local features, with $\alpha \in [10^{-3}, 10]$, equidistant in natural logarithm scale. The net weights are initialized from a kaiming normal initialization and bias are all set to zero. In this case only the AdamW optmizer is employed, with a cyclic learning rate scheduler. A minibatch size of 50 is employed and the training proceeds for as long as 2M epochs.

The parameters of the learning rate scheduler, as implemented in PyTorch are presented in Table~\ref{tab:lr_parameters}. Any other parameter is left to the default configuration.
\begin{table}[hbtp]
	\centering
	\begin{tabular}{c | c}
		scheduler & CyclicLR\\
		mode & triangular2\\
		base\_lr & $10^{-5}$\\
		max\_lr & $10^{-2}$ \\
		step\_size\_up & 1000\\
		cycle\_momentum & False
	\end{tabular}
	\label{tab:lr_parameters}
\end{table}